\documentclass[usenatbib,useAMS]{mnras}

\usepackage{graphics,epsf}
\usepackage{amsmath}                
\usepackage{amsfonts}               
\usepackage{amssymb}                
\usepackage{epsfig}                 
\usepackage{graphicx}

\usepackage{xcolor}
\definecolor{redak}{rgb}{0.8,0.05,0.05}

\def \kms{~\rm{km~s^{-1}}}

\def \msyr{~\rm{M_{\odot}}~\rm{yr^{-1}}}
\def \cm{~\rm{cm}}

\def \K{~\rm{K}}

\def \kpc{~\rm{kpc}}
\def \etc{$\eta$~Car~}
\def \days{~\rm{days}}

\def \rmModot{~\rm{M_{\sun}}}
\def \rmRodot{~\rm{R_{\sun}}}
\def \rmLodot{~\rm{L_{\sun}}}

\title[Accretion in HD 166734]{Wind collision and accretion simulations of the massive binary system HD 166734}

\author[A. Kashi]{Amit Kashi$^{1}$
\thanks{E-mail: \href{mailto:kashi@ariel.ac.il}{kashi@ariel.ac.il}}\\
$^{1}$Department of Physics, Ariel University, Ariel, POB 3, 4070000, Israel\\
}

\date{Accepted XXX. Received YYY; in original form ZZZ}

\pubyear{2020}

\begin{document}
\label{firstpage}
\pagerange{\pageref{firstpage}--\pageref{lastpage}}
\maketitle

\begin{abstract}
We run hydrodynamic simulations which follow the colliding winds structure of the massive binary system HD~166734 along its binary orbit, and show that close to periastron passage the secondary wind is suppressed and the secondary accretes mass from the primary wind.
The system consists two blue supergiants with masses of $M_1 \approx 39.5 \rmModot$ and $M_2 \approx 30.5 \rmModot$, on a $P \simeq 34.538 \days$ orbit with eccentricity of $e \approx 0.618$.
This close O-O binary with high eccentricity is observed through its orbit in the X-rays, where it shows an unusual long minimum close to periastron passage.
We use advanced simulations with wind acceleration and prescription treatment of accretion and simulate the entire orbit at high resolution that captures the instabilities in the winds. 
We find that the colliding wind structure is unstable even at apastron.
As the stars approach periastron passage the secondary wind is quenched by the primary wind and the accretion onto the secondary begins.
The accretion phase lasts for $\simeq 12 \days$, and the amount of accreted mass per cycle we obtain is $M_{\rm{acc}} \simeq 1.3 \cdot 10^{-8} \rmModot$.
The accretion phase can account for the observed decline in X-ray emission from the system.
\end{abstract}

\begin{keywords}
stars: individual: HD~166734 --- accretion, accretion discs --- stars: winds, outflows --- (stars:) binaries: general
\end{keywords}

\section{INTRODUCTION}
\label{sec:intro}

Massive stars have rich physics with special phenomena that are not seen in low mass stars, such as very strong mass loss and rapid transition between evolutionary stages.
A large fraction of massive stars reside in systems of binaries \citep[e.g.,][]{Sanaetal2012}.
As massive stars tend to have strong winds,
especially late in their evolution,
the massive binaries often present colliding winds that undergo shocks and emit X-rays.
The colliding winds may be stable or change quickly for some part of the stellar evolution.
In many cases the winds experience instabilities and deviate from their smooth conical shape predicted by classical theory and simulations \citep{Stevensetal1992, Dganietal1993, Dganietal1995, Dgani1993, DganiSoker1994, Vishniac1994, Stevens1995, WalderFolini1995, WalderFolini1996}.

If the binary is eccentric, the colliding wind structure changes its properties along the orbit.
In extreme cases, this change may lead to accretion of material onto the star with the wind that has lower momentum.
This process was predicted and developed theoretically for the most massive binary system known to date, \etc \citep{Soker2005a, Soker2005b, Soker2007, Akashietal2006, KashiSoker2009a, KashiSoker2009b}. This system has an eccentricity of 0.9, and a $\sim 1$--$2$ months of non-zero X-ray minimum. At about the same time of the minimum, a strong variability is observed in many spectral lines (e.g., \citealt{Daminelietal2008, Corcoranetal2015, Mehneretal2015, Hamaguchietal2016}, and references therein).
The accretion in \etc was obtained in high resolution numerical simulations \citep{Akashietal2013, Kashi2017}.
\cite{Kashi2019} extended this work to include the response of the secondary star to accretion, and obtained more accurate accretion rates.

Here we shall discuss a colliding winds binary system with two unevolved stars. Though less common than the WR-O binaries, this O-O binary is interesting due to its high eccentricity and momentum ratio.
The discussed system, HD~166734 (also known as V411~Ser) is composed of two massive blue supergiants, an O7.5If primary with mass $M_1= 39.5 \pm 6 \rmModot$ and an O9I(f) secondary with mass $M_2= 33.5 \pm 5 \rmModot$ \citep{Mahyetal2017,Nazeetal2017}.
The stars have an eccentric orbit ($e=0.618 \pm 0.02$) with a period $P \simeq 34.538 \days$, and have colliding winds that emit x-rays, with a minimum occurring close to periastron passage \citep{Mahyetal2017,Nazeetal2017}.

Being as bright as 8.3 mag, it already appears in the catalogue of \cite{Merrilletal1933}  (where it was classified as having a B8ne$\alpha$ spectrum).
HD~166734 was reported as an O8f star in the catalogue of \cite{Morganetal1955}, \cite{ContiAlschuler1971} classified it as an 07.5If star, and \cite{Fernie1972} as O9If star. 
All these classifications are of course predate the discovery of the binarity of the system, though it was known that the star is a variable.
The uniqueness of the system was noticed by \cite{EbbetsConti1978} who already gave masses estimates that are still withing the error-bars of the most modern models (that we discuss below).
The eccentricity they found, fitting an orbital solution to line velocities is 0.3. This estimate was later updated to 0.46 by \cite{Contietal1980}.
One of the comments of \cite{Contietal1980} was that the observed orbital eccentricity would not support a suggestion
that appreciable mass exchange has occurred in the
system. The mass exchange, which we find that does occur, is the main topic of the present paper.
\cite{Contietal1980} have also made very good estimates for the system mass loss rate, $\approx 10^{-5} \msyr$, which they inferred from the H$\alpha$ emission line strength and the formulations of \cite{KleinCastor1978}.
Another contribution of \cite{Contietal1980} was estimating the distance to the system, for which they obtained $2.3 \kpc$. This estimate was confirmed by \cite{Mahyetal2017}.
Periodic decreases in the system brightness, interpreted as
eclipses, were reported by \cite{OteroWils2005}, but only one per orbit was observed.
There is no known visual tertiary star around the central binary \citep{Sanaetal2014}.

\cite{Mahyetal2017} provided modern observations of the system, with many spectroscopic and photometric observations that covered the entire orbit thoroughly \citep[see also][]{Gossetetal2017}.
Their radial velocity measurements lead to a high accuracy orbital solution (some of the parameters are mentioned above).
The V,R, and I bands behave quite similar. They all show an increase of $\sim0.03$ mag peaking at phase $\sim - 0.01$ (phases are relative to the periastron epoch solution they found), and then a symmetrical decline of $\sim - 0.2$ mag, between phases $\sim 0$ and $\sim 0.04$ \citep{Mahyetal2017}.
They also ran a \texttt{CMFGEN} model \citep{HillierMiller1998} and compared it to stellar evolution model and determined the present stellar parameters as well as the initial masses of the stars.

\cite{Nazeetal2017} took X-ray observations of HD~166734 using Swift and XMM-Newton.
The system parameters they derived are close to the ones derived by \cite{Mahyetal2017}, and the differences are insignificant.
They found that the X-ray emission
strength varies along the orbit by about one order of magnitude. The x-ray light curve shows an increase in intensity, followed by a sharp decrease to a long minimum state lasting about $\Delta \phi \simeq 0.1$, from $\phi \simeq -0.01$ to $\phi \simeq 0.08$.

\cite{Nazeetal2017} found that during the minimum, the 
X-ray to bolometric luminosity is $\simeq 10^{-6.85}$, which is very close to the value observed in O-stars. From that they concluded that the wind collision structure does not contribute to the X-ray luminosity during the minimum.
They explored a number of scenarios, and found that the most likely one is that the colliding winds structure is totally (or almost totally) destroyed near periastron passage, and therefore there is no X-ray source.
They commented that this possibility is supported by the disappearance of the H$\alpha$ emission line that is
detected by \cite{Mahyetal2017} at periastron.
As the X-ray flux is not proportional to the inverse of the distance between the stars, as observed in other systems, \cite{Nazeetal2017} concluded that the wind collision in HD~166734 is non-adiabatic.

The stellar luminosities of the two stars, ${\log(L_1/\rmLodot)= 5.84 \pm 0.1}$ and ${\log(L_2/\rmLodot)= 5.73 \pm 0.1}$ \cite{Mahyetal2017, HigginsVink2018}, are larger than expected from standard stellar evolution models.
This is a part from a long lasting puzzle, where there is of a factor of 2 found between spectroscopic mass estimates and evolutionary mass estimates, known as the O-stars mass discrepancy problem \citep{Herreroetal1992},
A solution to this problem was obtained by \cite{WeidnerVink2010}, who came up with a novel new calibration of O-star spectral types.
The uniqueness of HD~166734 and the discrepancy observed in this system, lead \cite{HigginsVink2019} to model the stars with alternative parameters, following the method of \cite{WeidnerVink2010}.
\cite{HigginsVink2019} used the \texttt{MESA} stellar evolution code to create a calibrated grid of rotating star models, and found that the properties for the two stars in HD~166734 can be obtained with enhanced mixing by rotation and larger overshooting coefficient.
Their method also demonstrated that it is very useful to plot stellar evolution tracks on the mass-luminosity plane, in addition to the HR-diagram.

In this paper we use a high resolution hydrodynamic simulation to follow the structure of the colliding winds of HD~166734 across its entire orbit (section \ref{sec:sim}), and show that accretion onto the secondary close to periastron passage takes place (section \ref{sec:result}) and account for the X-ray minimum.
Our summary and discussion are given in section \ref{sec:summary}.

\section{Modeling Accretion}
\label{sec:sim}

\subsection{The Bondi-Holye-Lyttleton Accretion Rate}
\label{sec:sim:BHL}

We calculate the accretion rate according to the Bondi-Holye-Lyttleton (BHL) accretion theory
(\citealt{HoyleLyttleton1939, BondiHoyle1944}, see also a review by \citealt{Edgar2004}).
Differently from the classical theory, we here have more considerations that need to be addressed:
\begin{enumerate}
    \item Accretion is not from a constant density medium, but rather from wind that originates from a star and has a radial profile.
    \item The system is eccentric, therefore at different times that density of the wind at the position of the secondary is different, hence the accretion radius, and consequently the accretion rate change with time.
    \item The orbital speed creates another asymmetry in the accretion problem, as the relative velocity between the secondary and the primary wind changes with time. 
\end{enumerate}
All these considerations can be taken into account, as performed for the binary system $\eta$~Car in \cite{KashiSoker2009b}.

We mark $v_r$ as the radial component of the orbital velocity ($v_r$ is negative when the two stars approach each other) and $v_\theta$ as the tangential component of the orbital velocity (see figure 1 of \citealt{Soker2005b}).
The radial (along the line joining the two stars) component of the
relative velocity between the secondary star and the primary wind is
$v_1-v_r$.
The total relative speed between the secondary and the primary's wind is
\begin{equation}
v_{\rm wind1} = \sqrt{(v_1-v_r)^2 + v_\theta^2 }.
\label{vwind1}
\end{equation}
We then calculate the accretion radius
\begin{equation}
R_{\rm{BHL}}(\theta)=\frac{2GM_2}{v_{\rm{wind1}}^2} \label{bondi}.
\end{equation}
Moreover, the accretion plane (to be used with this accretion radius to form a circle) is perpendicular to the orbital plane
and rotated at an angle $\alpha_{\rm{wind1}}$ from the line connecting
the two stars, away from the primary, where
\begin{equation}
\tan(\alpha_{\rm{wind1}})=\frac{v_\theta}{v_1-v_r} \label{alphawind1}.
\end{equation}
We take the primary wind considering the acceleration by radiation according to a $\beta=1$ acceleration profile (further discussed in the following sections).
We slice the accretion circle (with an accretion radius $R_{\rm{acc}}=R_{\rm{BHL}}$) both radially and azimuthally into fine arcs. At the position of each of these arcs there is a different value of density. Between two time steps the amount of material that each of these arcs (together forming a thin cylinder) collects is summed, to obtain the mass accretion rate of this time-step.

\begin{figure}
\includegraphics[trim= 6cm 2.0cm 5.0cm 2.0cm,clip=true,width=0.99\columnwidth]{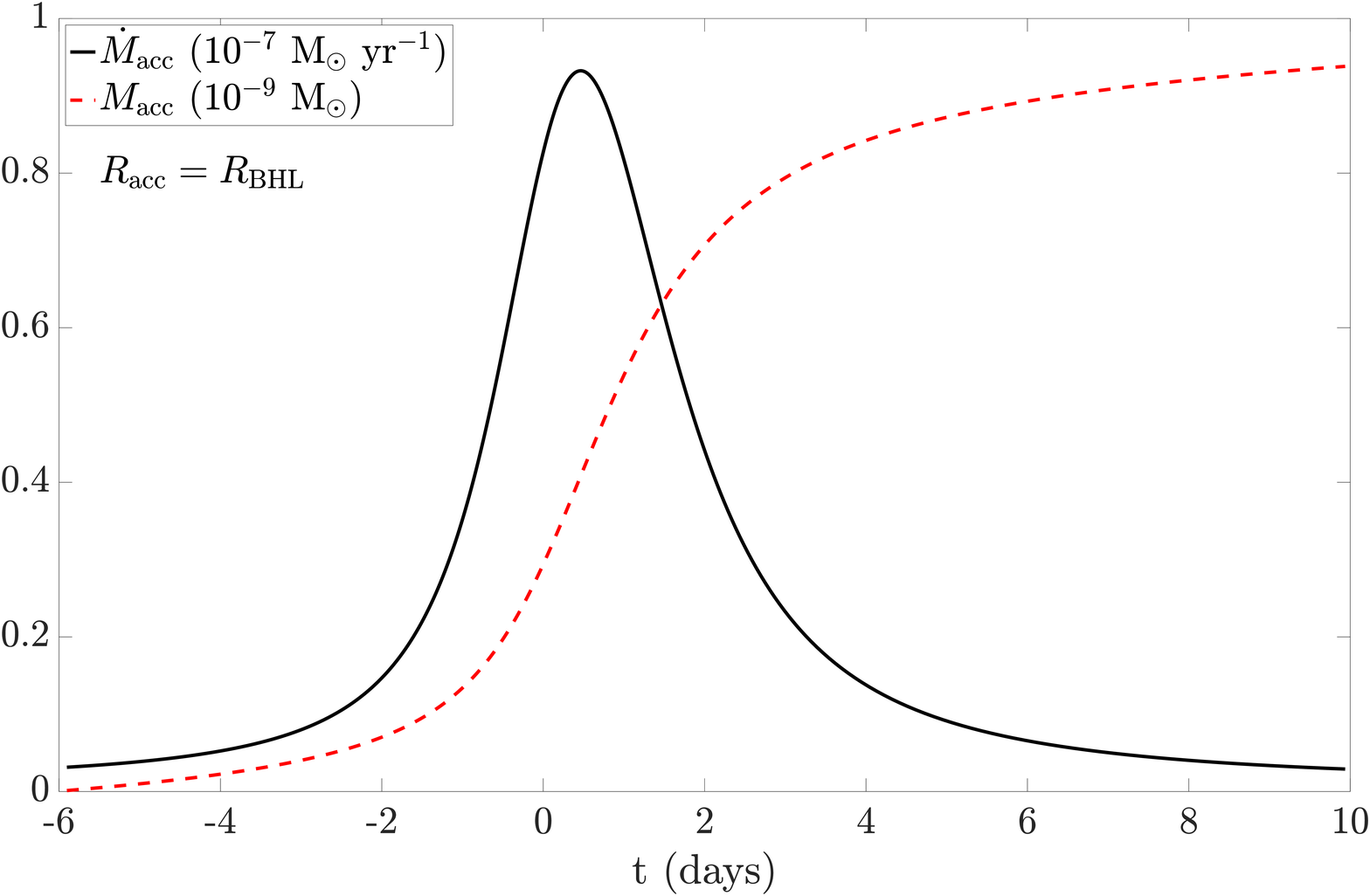} 
\includegraphics[trim= 6cm 0.0cm 5.0cm 2.0cm,clip=true,width=0.99\columnwidth]{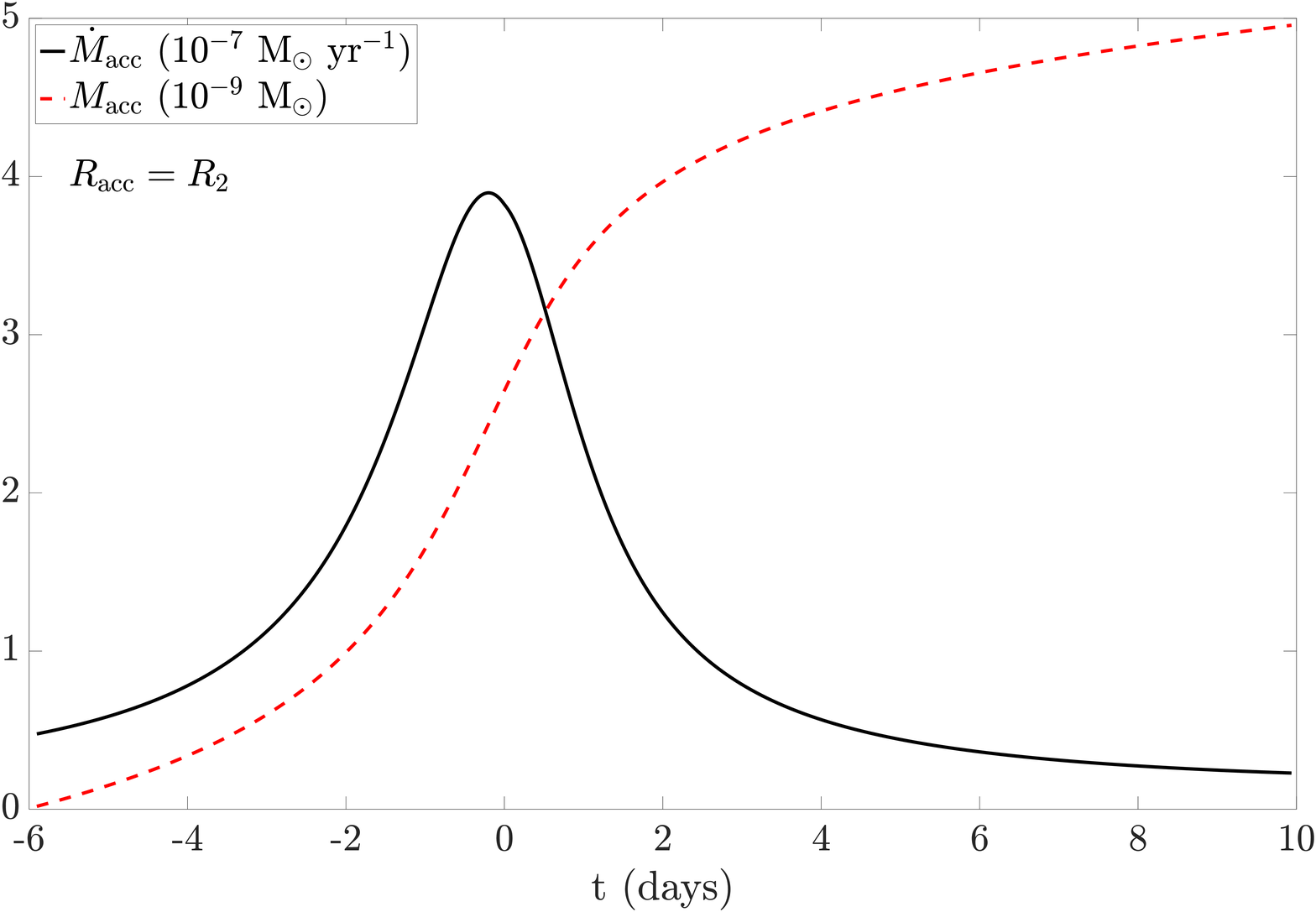} 
\caption{
\textit{Upper panel:} Computation results of the BHL accretion rate and accumulated accreted mass across one periastron passage.
The calculation takes into account additional effects, mainly the orbital velocity and the density profile of the primary wind (see text).
\textit{Lower panel:} The same calculation but taking the secondary radius $R_2$ as the accretion radius (note the different scale on the ordinate).
The BHL accretion radius is found to be smaller than the stellar radius, resulting in an underestimate of the accretion rate, and total accreted mass.
}
\label{fig:BHL}
\end{figure}

The upper panel of Figure \ref{fig:BHL} shows the accretion rate with a peak close to periastron passage, but a little after due to the asymmetry of the problem resulted by the relative velocity of the primary wind and the orbital motion. It also shows the total mass accreted in one periastron passage.
Since $R_{\rm{BHL}} < R_2$ for every $t$, our calculation clearly underestimates the accretion rate. We therefore repeat the same calculation, but with $R_{\rm{BHL}} = R_2$. The results are shown in the lower panel of Figure \ref{fig:BHL}.
We can see that the accreted mass is a few times larger, and the duration of accretion is quite similar in both cases, though the asymmetry is more evident in the case where the BHL accretion radius is used.
%

The final result of this exercise gives us an estimate for the amount of accreted mass under controlled conditions (i.e., no instabilities in the winds, smooth wind acceleration profile), $M_{\rm{acc}} \simeq 5 \cdot 10^{-9} \rmModot$.

\subsection{Effects of wind acceleration profile}
\label{sec:sim:beta}

The value we use for the acceleration profile of the winds, $\beta = 0.5$ which is the value originally derived for a point source according to the the CAK model \cite{CAK1975}.
\cite{GroenewegenLamers1991} found different values, in the range $\beta \simeq 0.5$--$1$ fit for each O-stars.
According to the modified CAK theory developed by \citep{Pauldrachetal1986}, and following papers analyzing additional effects \citep[e.g.,][]{Kudritzkietal1989} that became known as the Munich models, O-stars have wind profiles that obey $\beta \approx 0.8$--$1.0$ rather than $0.5$.
The modern models of \cite{MullerVink2008}, that solve analytically the velocity structure for any mass outflow or inflow situation through use of the Lambert W-function
also favoured $\beta \approx 0.8$.
The value $\beta \approx 0.8$ became quite accepted in the literature \citep{Pulsetal2008,Vink2015}.
On the other hand \cite{Guo2010} developed models for stellar winds and found that for O-stars $\beta \simeq 0.5$--$0.6$ fit the observations of X-rays from the stellar winds (from the individual winds themselves, not from the winds collision), since they have a large shock jump near the stellar surface.

The position of the colliding wind structure (to be precise, the contact discontinuity between the winds) along the line connecting the two stars is located where the ram pressure of the two winds is equal \citep[e.g.,][]{ShoreBrown1988, Stevensetal1992, Gayleyetal1997}.
This is a first-order estimate, neglecting radiation effects, instabilities and orbital motion.
Of course, there can theoretically be more than one solution, and the physical one is the one between the two peaks of the ram pressure lines
(see figure 2 in \citep{Stevensetal1992}).
In our case, where there are two O-stars with appreciable wind acceleration zone (in contrast to WR-O binaries where the small radius of the WR star allows neglecting its acceleration in most cases), the balance is obtained when
\begin{equation}
\frac{\dot{M}_1 v_{1,\infty}(1-R_1/(d-r))^\beta}{4\pi (d-r)^2} =
\frac{\dot{M}_2 v_{2,\infty}(1-R_2/r)^\beta}{4\pi r^2},
\label{eq:ram}
\end{equation}
where $r$ is the distance from the center of the secondary, and $d$ is the binary separation.
The left hand side is the ram pressure of the primary wind, which we mark $P_1$, and the right hand side is the ram pressure of the primary wind, which we mark $P_2$.

\begin{figure*}
\includegraphics[trim= 0.0cm 0.0cm 0.0cm 0.0cm,clip=true,width=0.99\textwidth]{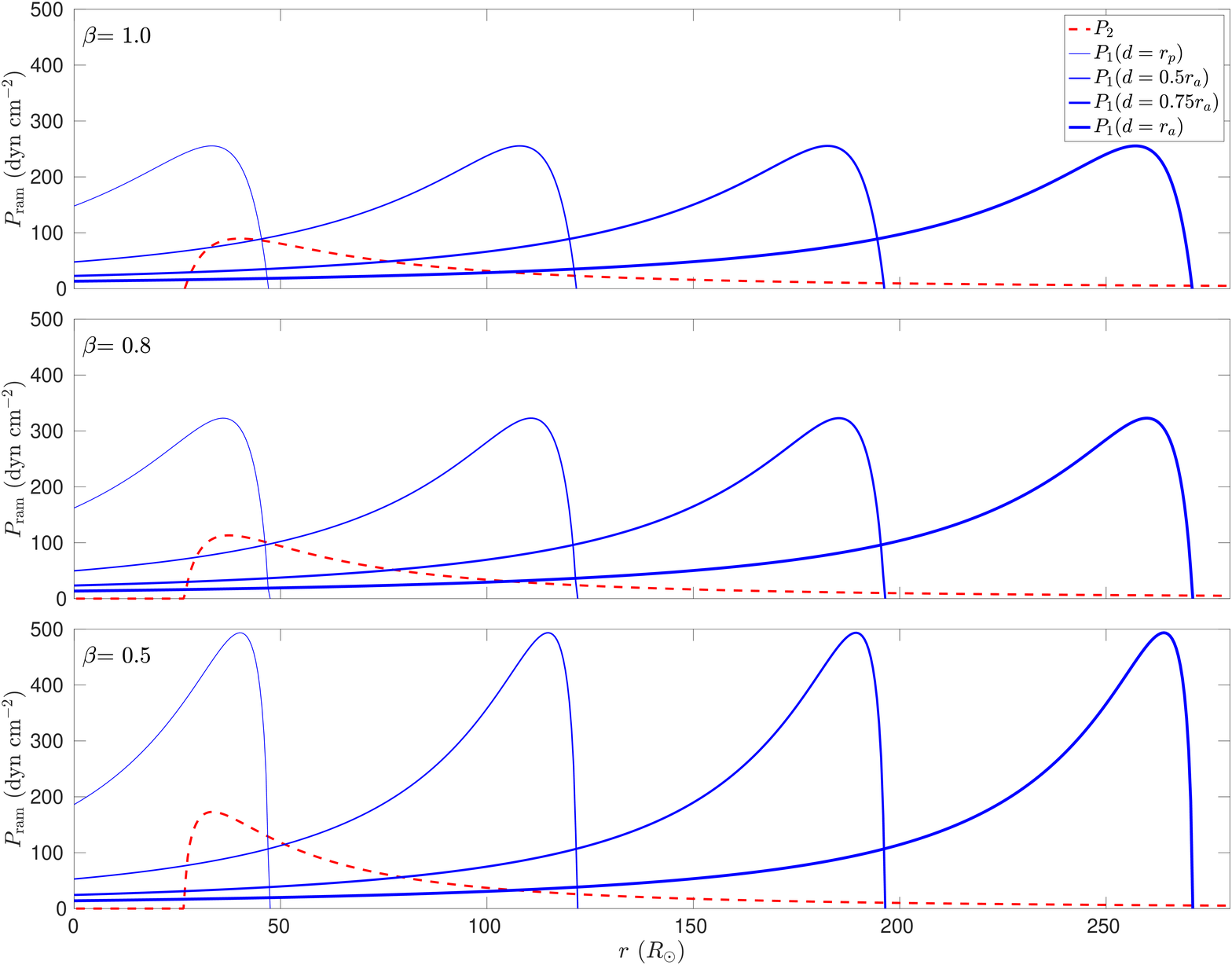} 
\caption{
The theoretical position of the colliding winds structure of HD~166734, along the line connecting the two stars.
We plot three different values of acceleration parameter $\beta$, and four different positions of the primary along its orbit ($r_a$, $r_p$ are the apastron and periastron distances, respectively).
The point where the two ram pressures are equal is the theoretical point where the contact discontinuity will be located.
In our simulations we use $\beta=0.5$ for the acceleration profile of both winds.
It is expected that a higher value of $\beta$ will lead to higher accretion rate and duration (see text).
}
\label{fig:beta}
\end{figure*}

Figure \ref{fig:beta} shows the ram pressure of the two winds. The upper, middle and lower panel are for $\beta = 1.0, 0.8 ,0.5$, respectively.
Both stars have the same $\beta$ index for each of the panels.
The ram pressure of the secondary is shown by the red-dashed line. On each panel (for each value of $\beta$, the ram pressure of the primary is plotted for 4 positions, from apastron to periastron.
Note that the ram pressure of the primary wind is not plotted only on the side of the primary facing the secondary, for clarity.
%

We can see from Figure \ref{fig:beta} that:
(1) At apastron and close to apastron, the colliding wind structure is in the same place, regardless of the value of $\beta$, because the winds reach terminal velocities for all values in all cases.
(2) At periastron distance, and close to periastron, there is no solution for the colliding wind structure.
(3) For higher value of $\beta$, the solution does not exists for a longer binary separation $r$, namely for a longer duration.

It is therefore expected that a higher value of $\beta$ will result in pushing the colliding winds structure towards the secondary earlier before periastron passage, and in turn will result the destruction of the colliding wind structure that lead to the beginning of the accretion phase. Similarly, for higher $\beta$ the accretion phase will end later.
Detailed calculations also showed that higher $\beta$ leads to higher mass flux, and in turn to higher accreted mass \citep{KashiSoker2009b}.
Therefore, higher $\beta$ will generally lead to more accretion of mass.
Therefore, if for the low value of $\beta=0.5$ we use, accretion will be obtained (and, as we show below, it is obtained), it will let alone be obtained for a higher value of $\beta$, and at a higher rate.

\subsection{The Numerical Simulation}
\label{sec:sim:Numerical}

We examine a model in which the colliding winds are accreted onto the secondary as the system undergoes periastron passage, and affect the X-ray observations of HD~166734. The idea behind the model is that the colliding wind structure does not exist close to periastron passage, and as it is the major contributor to the X-ray emission, the X-ray luminosity drops.
We emphasize that the simulation does not \textit{assume} accretion, but rather accretion, as we show below, is a \textit{result} of the simulation.

We use version 4.5 of the hydrodynamic code \texttt{FLASH}, originally described by \cite{Fryxell2000}.
Our 3D Cartesian grid extends over
$x\in[-180,360], y\in[-180,180], z\in[-90:90] \rmRodot$, including the entire binary orbit and beyond.
The secondary is located at $(x,y,z)=(0,0,0)$, and the primary is on a Keplerian orbit.
To solve the hydrodynamic equations we use the \texttt{FLASH} version of the split piece-wise parabolic method (PPM) solver \citep{ColellaWoodward1984}.
The code includes radiative cooling based on \cite{SutherlandDopita1993} and a prescription that describes the accretion.
The details regarding the modifications and enhancements we did to the code are described in \cite{Kashi2017} and \cite{Kashi2019}.
The two stars are treated as solid spheres. They are not part of the simulations and their hydrodynamic properties are not computed by the code.
The initial conditions are set in the way that the primary wind is present everywhere except in a sphere around the secondary.
The gas in our simulation follows an ideal-gas equation of state with adiabatic index $\gamma=5/3$.
The radiative cooling in the code, however, may reduce this value of  $\gamma$ according to the radiative energy losses.
The two winds are being ejected from the stars spherically, and accelerated according to the model described below.
Every time step the winds are accelerated unless accretion takes place. Material which is not originating in the accreting star is not being accelerated by its radiation.
The star are left stationary to relax and reach a quasi-steady state of colliding winds before the orbital motion begins.

We adopt the parameters for the orbit and the stellar properties from \cite{Mahyetal2017}.
For the primary we take $M_1= 39.5 \rmModot$, $R_1= 27.5 \rmRodot$, $T_{\rm{eff},1}= 32\,000 \K$ for the mass, radius and effective temperature, respectively. Its wind terminal velocity is $v_{1,\infty}= 1386 \kms$, and the mass loss rate is $\dot{M}_1= 9.07 \cdot 10^{-6} \msyr$.
For the secondary we take $M_2= 33.5 \rmModot$, $R_2= 26.8 \rmRodot$, $T_{\rm{eff},2}= 30\,500 \K$, $v_{2,\infty}= 1331\kms$ and $\dot{M}_2= 3.02 \cdot 10^{-6} \msyr$.
Both stars eject wind, and we adopt line-driven wind according to a CAK model \citep{CAK1975} with $\beta=0.5$.

The orbital period is $P\simeq34.538 \days$, and the eccentricity is $e\simeq0.618$ \citep{Mahyetal2017}
As the mass loss and mass transfer are small over one orbit, the deviation from Keplerian orbit is very small.
Our initial conditions are set at apastron, namely $\simeq17.26 \days$ before periastron passage which we define as $t=0$.
We let the simulation run at a stationary state, namely with the two stars fixed at their apastron locations, for $10 \days$. This allows for forming the colliding winds structure.
We use a high resolution grid, with a cell size of ($\simeq 0.07 \rmRodot$).
This level of resolution allows resolving the instabilities in the colliding winds structure.
and allows to follow in great detail the gas accreted onto the secondary.
The winds are initially smooth, but as the wind collides the instabilities create clumps and filaments. As will be shown below, these self-developed clumps are large and significant.
Therefore, there is no need to artificially seed any clumps in the stellar winds.

At apastron the two stars are far enough so wind s accelerate to their terminal velocity before they collide and create the colliding wind structure (also know as the wind-wind collision zone).
As the stars approach periastron passage the smaller distance between each star and the location where the winds collide does not enable the winds to reach their terminal velocities, and they collide while they are still accelerating.
The wind parameters result in momentum ratio
\begin{equation}
    \eta=\frac{\dot{M}_1 v_1}{\dot{M}_2 v_2} = 3.13. 
    \label{eq:eta}
\end{equation}
This is the value close to apastron. However, $\eta$ obtains higher values close to periastron passage, where, as mentioned, the secondary wind does not reach its terminal velocity.

As \cite{Nazeetal2017} found that the wind collision in HD~166734 is non-adiabatic, we include radiative cooling, according to the method described in \cite{Kashi2017}.
As we discuss below, the cooling leads to instabilities in the colliding winds structure (see section \ref{sec:result}).

The code we used has the ability to treat accreted material, should accretion take place.
The algorithm we adopt for the response of the secondary star to accretion is described in \cite{Kashi2019}.
It allows accelerating the secondary wind even when accretion starts.
The code is able to account for directional accretion, to within resolution of individual cell on the sphere that constitutes the surface of the secondary.

\section{Results}
\label{sec:result}
%
\begin{figure*}
\includegraphics[trim= 0.0cm 0.0cm 0.0cm 0.0cm,clip=true,width=0.89\textwidth]{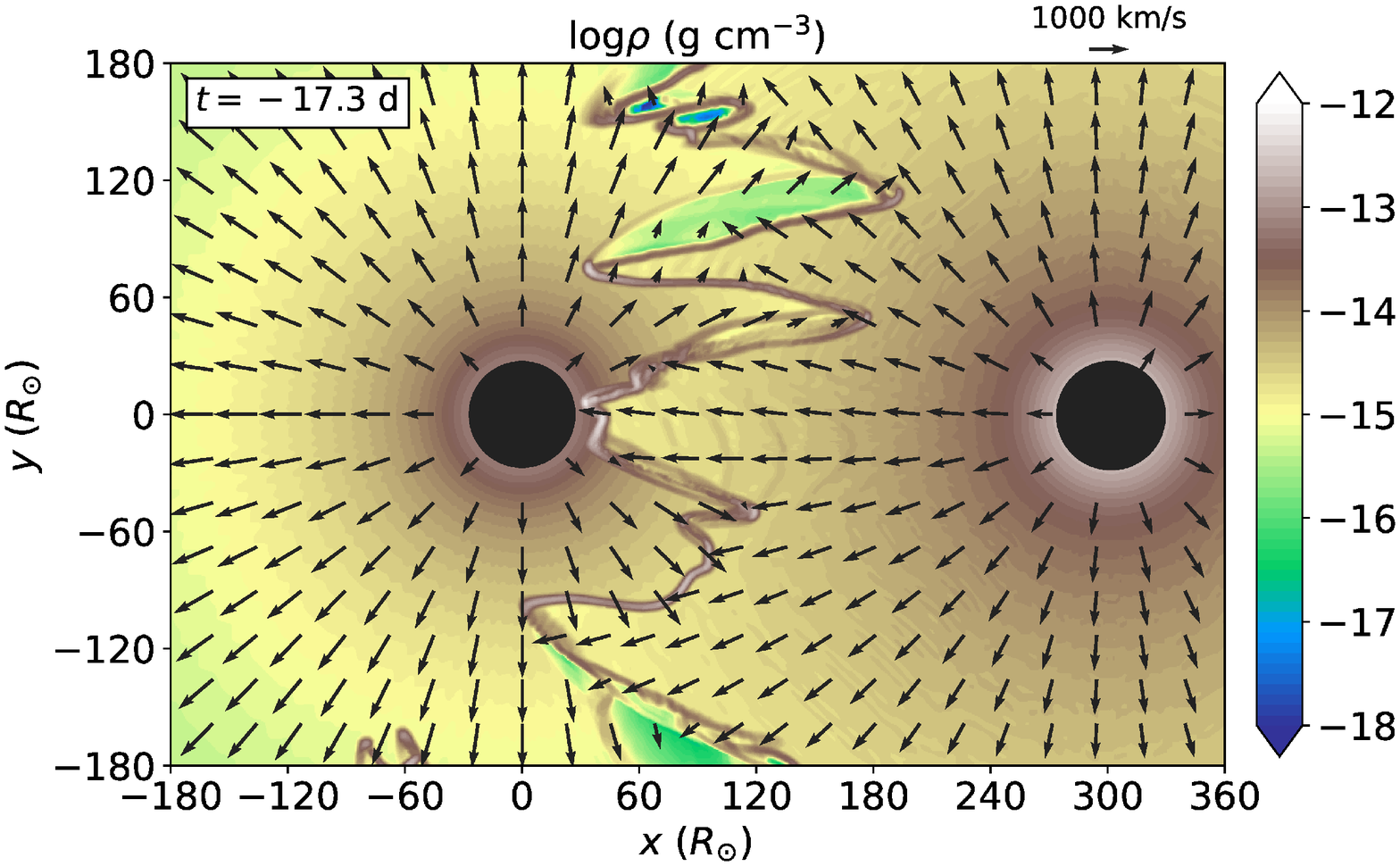} 
\includegraphics[trim= 2.5cm 0.0cm 0.5cm 1.5cm,clip=true,width=0.89\textwidth]{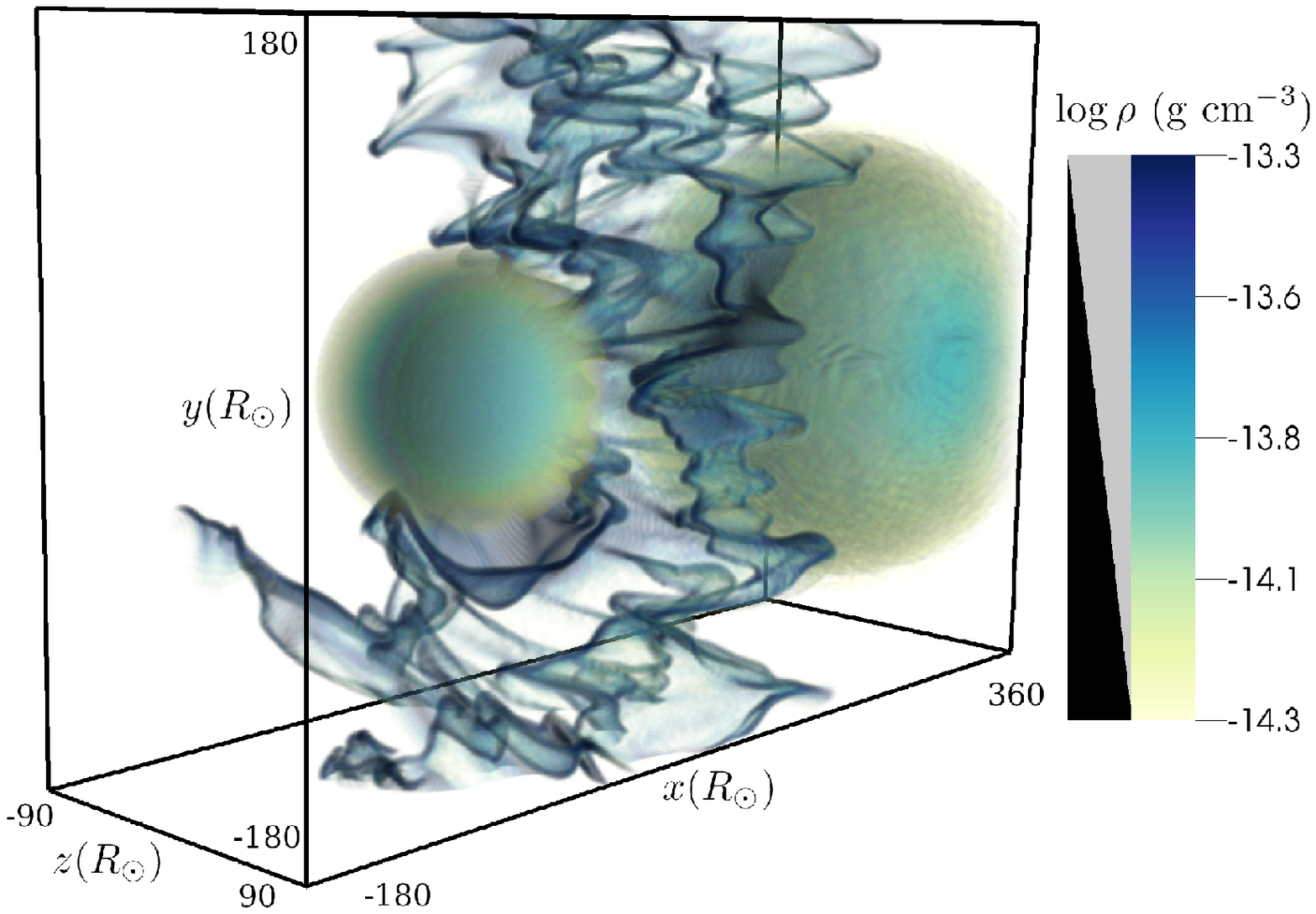} 
\caption{
\textit{Upper panel}: A density map with superimposed velocity vectors of our simulations at apastron ($17.27 \days$ before/after periastron passage). The slice is taken at $z=0$, namely on the orbital plane.
The two stars blow accelerating wind and kept stationary.
The winds collide when the two winds have essentially reached their terminal velocities.
We obtain an unstable wind solution with the NTSI present,
due to radiative cooling of both winds.
The instabilities changes the shape of the colliding wind structure, and creates layers of oblique shocks with ``fingers'' that penetrate deep into both sides of the undisturbed winds.
\textit{Lower Panel}: A 3D view of the density at apastron.
At the bottom of the figure the residuals of the gas that had engulfed the secondary when the primary was closer can still be seen.
}
\label{fig:apastron}
\end{figure*}
%
\begin{figure*}
\includegraphics[trim= 0.0cm 0.0cm 0.0cm 0.0cm,clip=true,width=0.99\textwidth]{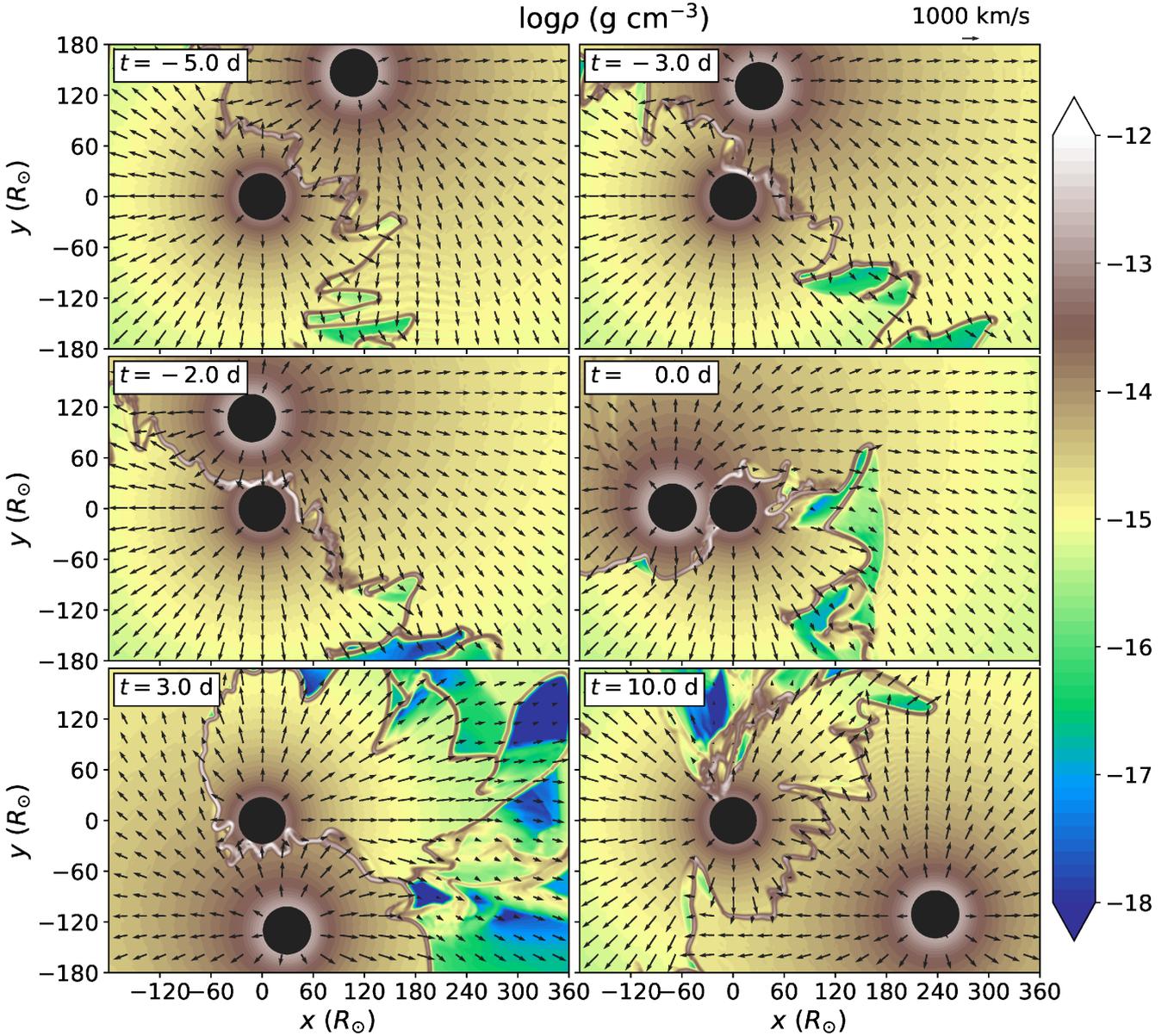} 
\caption{
A density map of our simulations, sliced on the orbital plane, at different times. The secondary is centered at the origin, orbited by the primary.
The two stars blow accelerating wind but as the system approaches periastron passage the secondary wind does not reach the terminal velocity  and the wind collision structure is pulled by the secondary gravity, approaching the secondary.
The wind solution is unstable with the NTSI present.
About two days before periastron passage accretion starts taking place.
As the two stars move away the secondary wind reinstates itself, and the situation return to the previous apastron-like state.}
\label{fig:snapshots}
\end{figure*}
%
\begin{figure*}
\includegraphics[trim= 2.50cm 0.0cm 0.5cm 1.0cm,clip=true,width=0.89\textwidth]{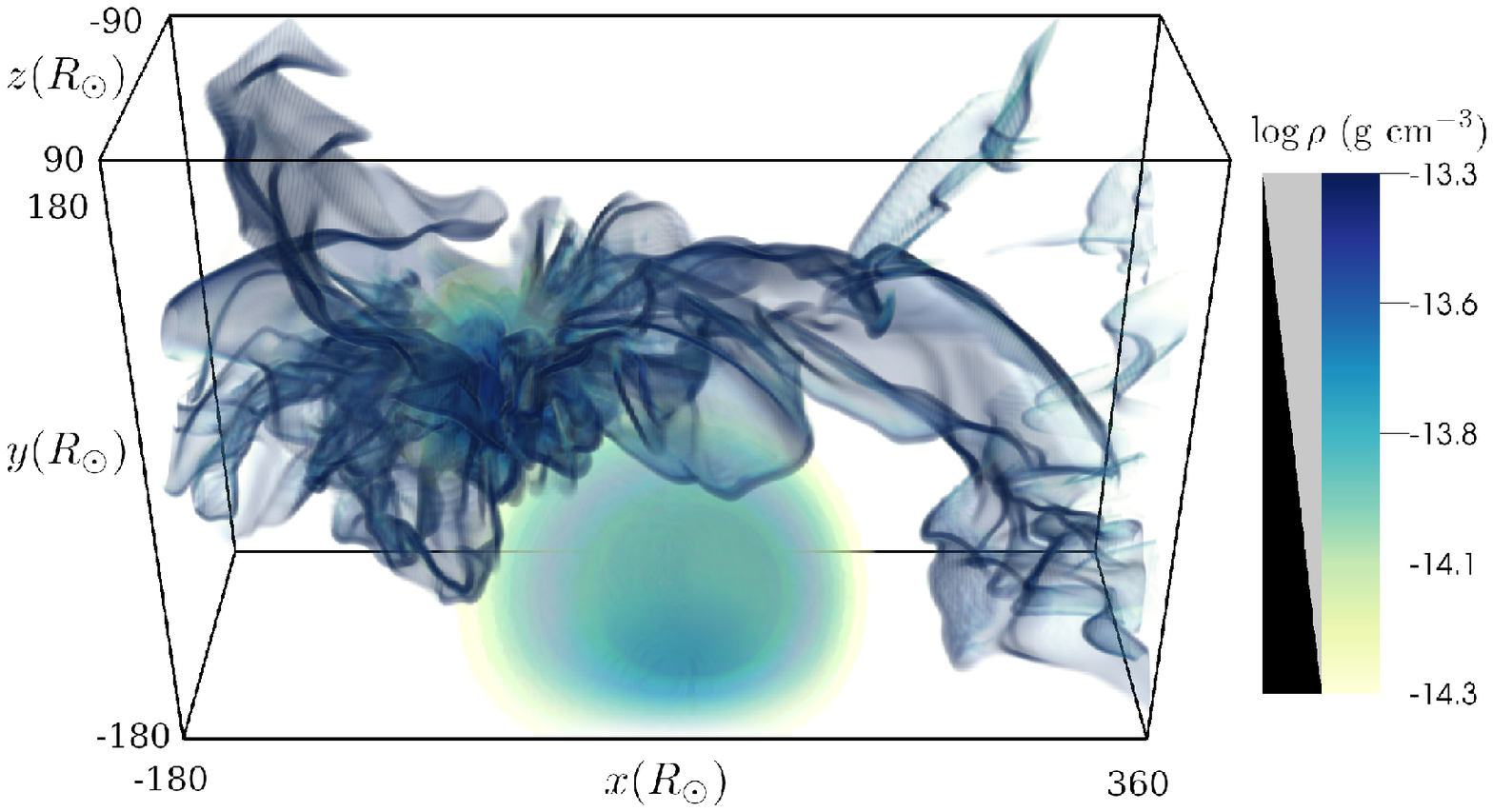} 
\caption{
A 3D view of the density during a time of a high accretion rate ($5 \days$ after periastron passage).
The general shape of the outer part of the colliding winds (that looks like the `$\backsim$' symbols) is the result of orbital motion of the winds.
The inner part is attracted to the secondary and accreted.
The accreted gas arrives to the secondary mostly from low latitudes.
}
\label{fig:acc3D}
\end{figure*}

The simulation shows that the wind is unstable even during the initial stage when the stars are stationary.
We can conclude that the instability is not the Kelvin-Helmholtz instability since (1) the winds are non-adiabatic, and (2) the instability shows even close to apastron, where the winds collide at almost their terminal velocity, which is comparable, and the Kelvin-Helmholtz instability requires velocity gradient.
However, even very small velocity gradient that causes a very small Kelvin-Helmholtz instability, or a small ripples as a result of the Rayleigh-Taylor instability can seed a significant non-linear thin-shell instability \cite[NTSI;][]{Vishniac1994}.
In addition, the transverse acceleration instability \citep{Dganietal1996a, Dganietal1996b} can take place when gas flows perpendicular to the curved surface of the shock.
The reason for having the instabilities, especially the NTSI, is mainly due to the rapid radiative cooling both winds undergo.
The properties of the NTSI, were simulated and thoroughly investigated by \cite{McLeodWhitworth2013}. They especially focused on the growth time of the perturbations at different wavelengths, and found a good match between simulation and theory.

We find that the NTSI modifies the shape of the colliding wind structure, and creates layers of oblique shocks with ``fingers'' that penetrate deep into both sides of the undisturbed winds (before the shocks), that have low density and cool to low temperatures.
Figure \ref{fig:apastron} shows the system at apastron.
The colliding wind structure did not relax to a bow shape as in adiabatic colliding winds, and we can see the fingers as a result of instabilities.
These fingers can reach very close to the secondary, but the secondary wind at apastron, and for most of the orbit is strong enough to prevent accretion.
Around the secondary we can see some of the residual gas from the colliding wind structure from the time when the stars were closer (between periastron passage and apastron).

The properties of the NTSI, were simulated and thoroughly investigated by \cite{McLeodWhitworth2013}. They especially focused on the growth time of the perturbations at different wavelengths, and found a good match between simulation and theory.
A similar effect was obtained in 2D simulations by \cite{Keeetal2014}, who also found that
the spatial dispersion of shock thermalization limits strong X-ray emission to the peaks and troughs of the fingers, and by that lowers the X-ray emission below the expected values from analytic radiative-shock models with stable colliding winds structure.

Figure \ref{fig:snapshots} shows a density map in the orbital plane for various times in the orbit.
As the stars approach periastron, the primary wind pushes the secondary wind to the extent that it almost stops, and only partially extends to the immediate vicinity of the secondary.
Approximately 2 days before periastron passage \textit{accretion of gas onto the secondary} begins.
The accretion starts taking place from different directions around the secondary, not only from the direction facing the primary (though it is stronger from that direction).

Figure \ref{fig:acc3D}
shows a 3D view of the density $5 \days$ after periastron passage, when the accretion rate is high.
The accreted gas arrives to the secondary mostly from low latitudes.
It arrives mostly radially from clumps that form and fall onto the secondary.

As mentioned earlier, the value of $\eta$ (equation \ref{eq:eta}) changes close to periastron passage.
The winds collide closer to the secondary, and then $v_2$ decreases and $\eta$ increases.
This process allows the gas to arrive more easily to the secondary.
The closer the colliding wind structure comes to the secondary, the weaker the resistance it encounters from the secondary wind.
In addition the gravity of the secondary works to pull the gas towards the secondary. This effect was isolated in the simulations of \cite{AkashiSoker2010}, and its significance was proven. 
All these together favour accretion, and can happen thanks to the high eccentricity that brings the stars very close.

The gravity of the secondary has an important role in attracting filaments and clumps of gas towards the secondary.
These filaments are the results of instability in the wind.
The accretion phase continues until $\sim 10$ days after periastron.
The secondary then recovers from the accretion phase, its wind is regained, and the colliding wind structure is then gradually restored.
We can see that the structure goes back to the wiggly shape it had on the previous cycle (as a result of the instabilities mentioned above).

\begin{figure}
\includegraphics[trim= 0.0cm 0.0cm 1.0cm 0.1cm,clip=true,width=0.99\columnwidth]{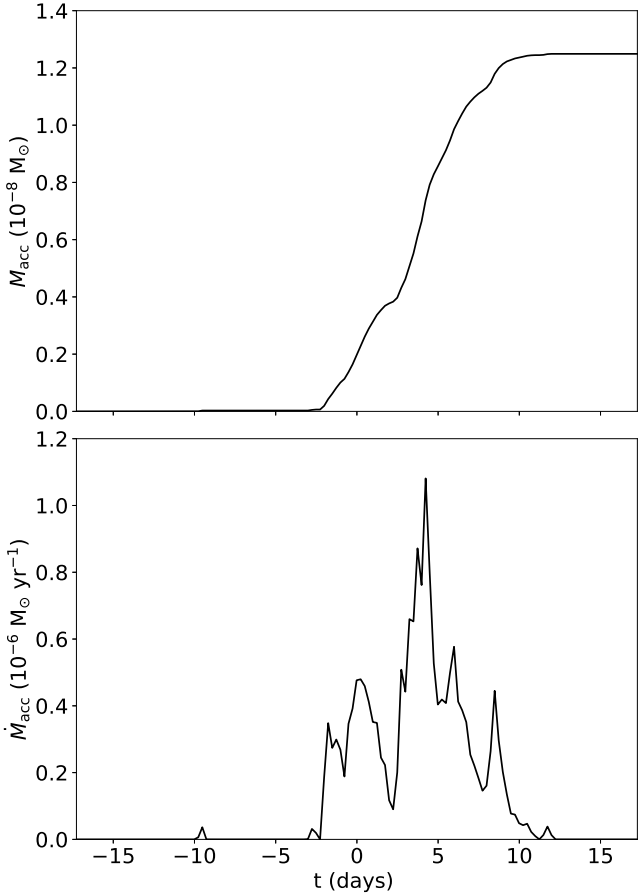} 
\caption{
Accretion rate (bottom panel) and accumulated accreted accreted mass onto the secondary (upper panel), as a function of time ($t=0$ denotes periastron passage).
About $\sim 2 \days$ before periastron passage accretion starts taking place, and lasts up to day $+10$ .
Overall, $\simeq 1.3 \cdot 10^{-8} \rmModot$ are accreted.
}
\label{fig:acc}
\end{figure}

We calculate the amount of mass accreted onto the secondary as a function of time. The result is shown in Figure \ref{fig:acc}. The bottom panel shows the accretion rate and the upper panel shows the integrated accreted mass. We can see that during the orbit the accreted mass reached $M_{\rm{acc}} \simeq 1.3 \cdot 10^{-8} \rmModot$, with a peak accretion rate of $\dot{M}_{\rm{acc}} \simeq 1.1 \cdot 10^{-6} \msyr$.
The mass is accreted in blobs, so peaks in the accretion rate indicate times where more blobs arrived.
The accretion is asymmetric relative to the time of periastron. This effect is also observed in \etc, and the reason is the relative velocity between the stars, that results in a lower relative wind velocity after periastron passage, and in turn a higher accretion rate (see section \ref{sec:sim:BHL} and Figure \ref{fig:BHL}).
%

We can see that the amount of accreted mass we obtained is larger, but on the same order of magnitude as the the BHL value.
The major source for the difference, is an effect that cannot be accounted for in the analytic computation -- the presence of colliding winds that develop instabilities.
As we see from the simulations, these instabilities result in episodes of high accretion rates.

Most of the X-ray emission comes from the post-shocked colliding winds.
Radiative cooling can significantly reduce the X-ray emission.
We calculate the X-ray emission from the binary system, taking the contribution of every cell $j$
\begin{equation}
L_{x,em,j}=\Lambda n_p n_e V
\label{eq:Lx_em}
\end{equation}
Where $\Lambda$ is the cooling function, that we adopt from \citealt{SutherlandDopita1993}, $V$ is the emitting volume, and $n_p$ and $n_e$ are respectively the proton and electron number densities.

We perform a detailed calculation for the absorption of X-rays in the winds and the colliding wind structure.
From every cell that gets hot enough and emits X-rays, we calculate the hydrogen column density $N_{H,j}$ along the line of sight towards the observer, situated in inclination angle $i=63^\circ$ and argument of periapsis $\omega=236.183^\circ$ \citep{Nazeetal2017}.
We then compute the absorption $\tau_j=\sigma N_{H,j}$, taking $\sigma=10^{-22} \cm^2$, an average value for the $0.5$--$10~\rm{KeV}$ range.
The transmitted part is then
\begin{equation}
L_{x,j}=L_{x,em,j} e^{-\tau_j}.
\label{eq:Lx}
\end{equation}
Then we sum all the cells to obtain the X-ray luminosity $L_x$.
The above procedure is performed in post-processing of the simulation, and as it is done cell-by-cell in high resolution it is computationally heavy by itself.

\begin{figure}
\includegraphics[trim= 1.0cm 0.0cm 1.5cm 1.0cm,clip=true,width=0.99\columnwidth]{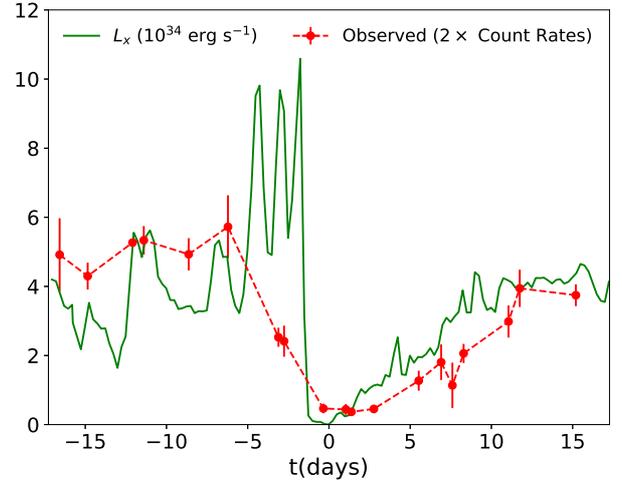} 
\caption{
The X-ray emission from our simulation (green line), and the $0.5$--$10 \rm{KeV}$ X-ray counts observed for HD~166734 (red dots connected with a dashed red line; \citealt{Nazeetal2017})
The X-ray emission decreases as accretion starts, and restored when accretion stops.
The observations and the simulation are qualitatively similar.
A few days before periastron there is a mismatch between them, due to lack of observations that probably miss the expected X-ray flares that are formed as a results of clumping.
}
\label{fig:xray1}
\end{figure}

\begin{figure}
\includegraphics[trim= 1.0cm 0.0cm 1.5cm 1.0cm,clip=true,width=0.99\columnwidth]{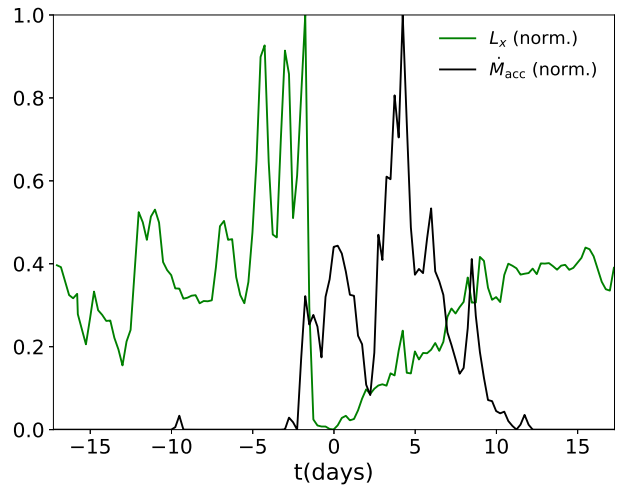} 
\caption{
The normalized X-ray emission (green line; from Figure \ref{fig:xray1}), and the normalized accretion rate (black; from Figure \ref{fig:acc}) from our simulation .
}
\label{fig:xray2}
\end{figure}

Figure \ref{fig:xray1} compares the X-ray luminosity obtained obtained from our simulation, and the the $0.5$--$10~\rm{KeV}$ X-ray emission reported by \cite{Nazeetal2017}.
The results are qualitatively similar, except of days $-5$--$-1$, namely shortly before periastron passage.
This is however a time of a gap in the observations.
We obtain during this time X-ray flares.
Such flares have been observed in other colliding wind systems, and are the results of clumps that form close to periastron passage \citep[e.g.,][]{MoffatCorcoran2009,Ramiaramanantsoaetal2019}.
There are only two very close observations, that have low count rates, during these $\approx 4$ days, and they could easily been taken coincidentally between flares.
Figure \ref{fig:xray2} shows the normalized X-ray emission, and the normalized accretion rate from our simulation.
It can be seen that the X-ray emission decreases as accretion starts, and restored when accretion stops.
%
%

\section{Summary and Discussion}
\label{sec:summary}

We presented high resolution hydrodynamic simulations of the binary system HD~166734, across its entire orbit.
The code simulates the hydrodynamics of the colliding winds at high resolution that allows instabilities, such as the non-linear thin shell instability (NTSI; \citealt{Vishniac1994}) seed and develop.
These instabilities are time variable during the orbital motion.
The high eccentricity of the system ($e\simeq0.618$) resulted in a fundamental difference between apastron and periastron.
About two days before periastron passage the primary wind starts being accreted onto the secondary, and lasts up to day $+10$ .
As the two stars move away the secondary wind reinstates itself, and the situation return to the way it was, with a colliding wind structure as before.
The close orbit of HD~166734 results in a major influence of the secondary gravity and accretion along a large fraction of the orbit.
The accretion we obtain is a result of the simulation, and was not assumed in any way.
Overall, $\simeq 1.3 \cdot 10^{-8} \rmModot$ are accreted every orbital cycle.
This is about $\simeq 2.5$ times larger the value from our analytic computation in section \ref{sec:sim:BHL}, suggesting that the analytic model, that requires negligible computational resources, can indeed give a reliable order of magnitude estimation.

We calculated the expected X-ray emission with a detailed post-processing computation. In this computation we used the orientation suggested for the system by \cite{Nazeetal2017}, and obtained qualitatively the same results.
Therefore, we consider the orientation obtained by \cite{Nazeetal2017} plausible.
We note however, that as other orientations were not tested we cannot rule out other possibilities  (as was demonstrated for \etc, e.g., \citealt{KashiSoker2009a}).

\cite{ParkinGosset2011} discuss the ``collapse'' of the wind-wind collision structure onto the O-star in the {WR-O} binary system WR~22.
It is in place to make a clear distinction between ``collapse'' and ``accretion''.
As \cite{ParkinGosset2011} write, during a collapse of the wind collision region onto the O star, the wind of the WR star will enter deep into the O star wind acceleration region, and may potentially collide against the O star photosphere. But they do not expect that this will lead to significant accretion and mass transfer as the WR star wind has too great a kinetic energy to become bound to the O star.

We consider material that reach the surface of the star as accreted. It is unclear why the kinetic energy should prevent the accretion.
If one uses the (scalar) Bernoulli equation to determine if material is bound or not, one has to remember that it is appropriate for velocity vectors moving away from the massive object.
But in the case where material reaches the stellar photosphere in close to radial velocities pointing towards the center of the star, the material will certainly be accreted.

As seen in Figure \ref{fig:acc3D}, gas was accreted onto the secondary mostly from equatorial regions.
As the accretion phase is relatively short, and the stellar wind of the secondary is strong, the material does not stay long enough to form an accretion disk.
Another reason is the large radius of the O-star, compared to e.g., a WR-star.
It is however possible that in other systems the angular momentum of the accreted material will form an accretion disk, and maybe launch jets or a polar outflow.

Accretion in massive binary stars is a process that may have a significant influence on the evolution of the stars, as the accreted massed can play a major role.
For the duration of a few million years of evolution, a few~$\cdot \rmModot$ can be transferred between the stars. The influence is significant both for the donor and the gainer. The effects of losing and especially gaining a lot of mass can be important, and should be further studied in stellar evolution codes.

\vspace{0.5cm}
We thank an anonymus referee for helpful comments.
We acknowledge PRACE for awarding us access to the national supercomputer Cray XC40 at the High Performance Computing Center Stuttgart (HLRS) under the grant number PrcPA4873/ipramkas.
Computations presented in this work were performed on the Hive computer cluster at the University of Haifa.
We acknowledge support from the R\&D authority, and the Chairman of the Department of Physics in Ariel University.

\label{lastpage}

\begin{thebibliography}{}

\bibitem[Akashi et al.(2013)]{Akashietal2013} Akashi, M.~S., Kashi, A., \& Soker, N.\ 2013, \na, 18, 23

\bibitem[Akashi \& Soker(2010)]{AkashiSoker2010} Akashi, M., \& Soker, N.\ 2010, arXiv e-prints, arXiv:1006.3333.

\bibitem[Akashi et al.(2006)]{Akashietal2006} Akashi, M., Soker, N., \& Behar, E.\ 2006, \apj, 644, 451

\bibitem[Bondi \& Hoyle(1944)]{BondiHoyle1944} Bondi, H., \& Hoyle, F.\ 1944, \mnras, 104, 273 

\bibitem[Castor, Abbott \& Klein(1975)]{CAK1975} Castor, J.~I., Abbott, D.~C., \& Klein, R.~I.\ 1975, \apj, 195, 157

\bibitem[Colella \& Woodward(1984)]{ColellaWoodward1984} Colella, P., \& Woodward, P.~R.\ 1984, Journal of Computational Physics, 54, 174 

\bibitem[Conti \& Alschuler(1971)]{ContiAlschuler1971} Conti, P.~S., \& Alschuler, W.~R.\ 1971, \apj, 170, 325.

\bibitem[Conti et al.(1980)]{Contietal1980} Conti, P.~S., Massey, P., Ebbets, D., et al.\ 1980, \apj, 238, 184

\bibitem[Corcoran et al.(2015)]{Corcoranetal2015} Corcoran, M.~F., Hamaguchi, K., Liburd, J.~K., et al.\ 2015, arXiv:1507.07961 

\bibitem[Damineli et al.(2008)]{Daminelietal2008} Damineli, A., Hillier, D.~J., Corcoran, M.~F., Stahl O., Groh J.~H., Arias J., Teodoro M., \& Morrell N.\ 2008b, \mnras, 386, 2330

\bibitem[Dgani(1993)]{Dgani1993} Dgani, R.\ 1993, \aap, 271, 527.

\bibitem[Dgani \& Soker(1994)]{DganiSoker1994} Dgani, R., \& Soker, N.\ 1994, \aap, 282, 54.

\bibitem[Dgani, Soker, \& Cadavid(1995)]{Dganietal1995} Dgani, R., Soker, N., \& Cadavid, M.~L.\ 1995, \aj, 110, 1894.

\bibitem[Dgani, Walder, \& Nussbaumer(1993)]{Dganietal1993} Dgani, R., Walder, R., \& Nussbaumer, H.\ 1993, \aap, 267, 155.

\bibitem[Dgani et al.(1996a)]{Dganietal1996a} Dgani, R., van Buren, D., \& Noriega-Crespo, A.\ 1996, \apj, 461, 372

\bibitem[Dgani et al.(1996b)]{Dganietal1996b} Dgani, R., van Buren, D., \& Noriega-Crespo, A.\ 1996, \apj, 461, 927 

\bibitem[Ebbets, \& Conti(1978)]{EbbetsConti1978} Ebbets, D.~C., \& Conti, P.~S.\ 1978, \baas 10, 631

\bibitem[Edgar(2004)]{Edgar2004} Edgar, R.\ 2004, \nar, 48, 843.

\bibitem[Fernie(1972)]{Fernie1972} Fernie, J.~D.\ 1972, \aj, 77, 150.

\bibitem[Fryxell et al.(2000)]{Fryxell2000} Fryxell, B., Olson, K., Ricker, P., et al.\ 2000, \apjs, 131, 273


\bibitem[Gayley et al.(1997)]{Gayleyetal1997} Gayley, K.~G., Owocki, S.~P., \& Cranmer, S.~R.\ 1997, \apj, 475, 786

\bibitem[Gosset et al.(2017)]{Gossetetal2017} Gosset, E., Mahy, L., Damerdji, Y., et al.\ 2017, The Lives and Death-throes of Massive Stars, 402

\bibitem[Groenewegen \& Lamers(1991)]{GroenewegenLamers1991} Groenewegen, M.~A.~T., \& Lamers, H.~J.~G.\ 1991, \aap, 243, 429.

\bibitem[Guo(2010)]{Guo2010} Guo, J.~H.\ 2010, \aap, 512, A50.

\bibitem[Hamaguchi et al.(2016)]{Hamaguchietal2016} Hamaguchi, K., Corcoran, M.~F., Gull, T.~R., et al.\ 2016, \apj, 817, 23 

\bibitem[Herrero et al.(1992)]{Herreroetal1992} Herrero, A., Kudritzki, R.~P., Vilchez, J.~M., et al.\ 1992, \aap, 261, 209.

\bibitem[Higgins, \& Vink(2018)]{HigginsVink2018} Higgins, E.~R., \& Vink, J.~S.\ 2018, arXiv e-prints, arXiv:1810.12924

\bibitem[Higgins, \& Vink(2019)]{HigginsVink2019} Higgins, E.~R., \& Vink, J.~S.\ 2019, \aap, 622, A50

\bibitem[Hillier \& Miller(1998)]{HillierMiller1998} Hillier, D.~J., \& Miller, D.~L.\ 1998, \apj, 496, 407.

\bibitem[Hoyle \& Lyttleton(1939)]{HoyleLyttleton1939} Hoyle, F., \& Lyttleton, R.~A.\ 1939, Proceedings of the Cambridge Philosophical Society, 35, 405 

\bibitem[Kashi(2017)]{Kashi2017} Kashi, A.\ 2017, \mnras, 464, 775

\bibitem[Kashi(2019)]{Kashi2019} Kashi, A.\ 2019, \mnras, 486, 926

\bibitem[Kashi \& Soker(2009a)]{KashiSoker2009a} Kashi, A., \& Soker, N.\ 2009a, \mnras, 397, 1426

\bibitem[Kashi \& Soker(2009b)]{KashiSoker2009b} Kashi, A., \& Soker, N.\ 2009b, \na, 14, 11

\bibitem[Kee, Owocki, \& ud-Doula(2014)]{Keeetal2014} Kee, N.~D., Owocki, S., \& ud-Doula, A.\ 2014, \mnras, 438, 3557.

\bibitem[Klein \& Castor(1978)]{KleinCastor1978} Klein, R.~I., \& Castor, J.~I.\ 1978, \apj, 220, 902

\bibitem[Kudritzki et al.(1989)]{Kudritzkietal1989} Kudritzki, R.~P., Pauldrach, A., Puls, J., \& Abbott, D.~C.\ 1989, \aap, 219, 205 

\bibitem[Mahy et al.(2017)]{Mahyetal2017} Mahy, L., Damerdji, Y., Gosset, E., et al.\ 2017, \aap, 607, A96

\bibitem[McLeod \& Whitworth(2013)]{McLeodWhitworth2013} McLeod, A.~D., \& Whitworth, A.~P.\ 2013, \mnras, 431, 710.

\bibitem[Mehner et al.(2015)]{Mehneretal2015} Mehner, A., Davidson, K., Humphreys, R.~M., et al.\ 2015, \aap, 578, A122

\bibitem[Merrill \& Burwell(1933)]{Merrilletal1933} Merrill, P.~W., \& Burwell, C.~G.\ 1933, \apj, 78, 87.

\bibitem[Moffat \& Corcoran(2009)]{MoffatCorcoran2009} Moffat, A.~F.~J., \& Corcoran, M.~F.\ 2009, \apj, 707, 693.


\bibitem[Morgan, Code, \& Whitford(1955)]{Morganetal1955} Morgan, W.~W., Code, A.~D., \& Whitford, A.~E.\ 1955, \apjs, 2, 41.

\bibitem[M{\"u}ller \& Vink(2008)]{MullerVink2008} M{\"u}ller, P.~E., \& Vink, J.~S.\ 2008, \aap, 492, 493.

\bibitem[Naz{\'e} et al.(2017)]{Nazeetal2017} Naz{\'e}, Y., Gosset, E., Mahy, L., et al.\ 2017, \aap, 607, A97

\bibitem[Otero \& Wils(2005)]{OteroWils2005} Otero, S.~A., \& Wils, P.\ 2005, Information Bulletin on Variable Stars, 5644, 1.

\bibitem[Parkin, \& Gosset(2011)]{ParkinGosset2011} Parkin, E.~R., \& Gosset, E.\ 2011, \aap, 530, A119

\bibitem[Pauldrach et al.(1986)]{Pauldrachetal1986} Pauldrach, A., Puls, J., \& Kudritzki, R.~P.\ 1986, \aap, 164, 86 

\bibitem[Puls et al.(2008)]{Pulsetal2008} Puls, J., Vink, J.~S., \& Najarro, F.\ 2008, \aapr, 16, 209.

\bibitem[Ramiaramanantsoa et al.(2019)]{Ramiaramanantsoaetal2019} Ramiaramanantsoa, T., Ignace, R., Moffat, A.~F.~J., et al.\ 2019, \mnras, 490, 5921.

\bibitem[Sana et al.(2012)]{Sanaetal2012} Sana, H., de Mink, S.~E., de Koter, A., et al.\ 2012, Science, 337, 444.

\bibitem[Sana et al.(2014)]{Sanaetal2014} Sana, H., Le Bouquin, J.-B., Lacour, S., et al.\ 2014, \apjs, 215, 15.

\bibitem[Shore \& Brown(1988)]{ShoreBrown1988} Shore, S.~N., \& Brown, D.~N.\ 1988, \apj, 334, 1021.


\bibitem[Soker(2005a)]{Soker2005a} Soker, N.\ 2005a, \apj, 619, 1064

\bibitem[Soker(2005b)]{Soker2005b} Soker, N.\ 2005b, \apj, 635, 540.

\bibitem[Soker(2007)]{Soker2007} Soker, N.\ 2007, \apj, 661, 482.

\bibitem[Stevens(1995)]{Stevens1995} Stevens, I.~R.\ 1995, Wolf-Rayet Stars: Binaries; Colliding Winds; Evolution, 163, 486.

\bibitem[Stevens et al.(1992)]{Stevensetal1992} Stevens, I.~R., Blondin, J.~M., \& Pollock, A.~M.~T.\ 1992, \apj, 386, 265.

\bibitem[Sutherland \& Dopita(1993)]{SutherlandDopita1993} Sutherland, R.~S., \& Dopita, M.~A.\ 1993, \apjs, 88, 253.


\bibitem[Vink(2015)]{Vink2015} Vink, J.~S.\ 2015, Very Massive Stars in the Local Universe:,
Astrophysics and Space Science Library, Volume 412. ISBN 978-3-319-09595-0. Springer International Publishing, Switzerland.


\bibitem[Vishniac(1994)]{Vishniac1994} Vishniac, E.~T.\ 1994, \apj, 428, 186.

\bibitem[Walder \& Folini(1995)]{WalderFolini1995} Walder, R., \& Folini, D.\ 1995, Wolf-Rayet Stars: Binaries; Colliding Winds; Evolution, 163, 525.

\bibitem[Walder \& Folini(1996)]{WalderFolini1996} Walder, R., \& Folini, D.\ 1996, \aap, 315, 265.

\bibitem[Weidner \& Vink(2010)]{WeidnerVink2010} Weidner, C., \& Vink, J.~S.\ 2010, \aap, 524, A98.

\end{thebibliography}
\end{document}